\documentclass[floats, nofootinbib, preprint]{revtex4-1}
\usepackage{bm,amsfonts, mathtools}
\usepackage{graphicx,color, wasysym}
\usepackage{textcomp}
\usepackage{amsmath,amssymb,latexsym,epsfig}

 \def\ulamek#1#2{\mbox{\normalfont$\frac{#1}{#2}$}}

\begin{document}

\makeatletter

\title{Exact and explicit evaluation of Br\'{e}zin-Hikami kernel}

\author{K.~G\'{o}rska}
\email{kasia_gorska@o2.pl}
\affiliation{H. Niewodnicza\'{n}ski Institute of Nuclear Physics, Polish Academy of Sciences, ul.Eljasza-Radzikowskiego 152, 
PL 31342 Krak\'{o}w, Poland}

\author{K.~A.~Penson}
\email{penson@lptl.jussieu.fr}
\affiliation{Laboratoire de Physique Th\'eorique de la Mati\`{e}re Condens\'{e}e,\\
Universit\'e Pierre et Marie Curie, CNRS UMR 7600\\
Tour 13 - 5i\`{e}me \'et., B.C. 121, 4 pl. Jussieu, F 75252 Paris Cedex 05, France\vspace{2mm}}

\def\ulamek#1#2{\mbox{\normalfont$\frac{#1}{#2}$}}

\begin{abstract}
We present exact and explicit form of the kernels $\hat{K}(x, y)$ appearing in the theory of energy correlations in the ensembles of Hermitian random matrices with Gaussian probability distribution, see E.~Br\'{e}zin and S.~Hikami, Phys. Rev. E 57 (1998) 4140 and E 58 (1998) 7176. In obtaining this result we have exploited the analogy with the method of producing exact forms of two-sided, symmetric L\'{e}vy stable laws, presented by us recently. This result is valid for arbitrary values of parameters in question. We furnish analytical and graphical representations of physical quantities calculated from $\hat{K}(x, y)$'s.
\end{abstract}

\maketitle

\section{Introduction}

We consider in this work certain extensions of results obtained by Br\'{e}zin and Hikami in \cite{EBrezin98}, hereafter referred to as BH. These authors have developed a theory of energy correlations in the ensembles of Hermitian random matrices characterized by Gaussian probability distribution. In the following we adopt the notation used by BH. In Sec. III of BH the Hamiltonian has a deterministic part $H_{0}$ with the spectrum consisting of two distinct degenerated levels and the random part $V$. This produces the singularity in the density of states $\rho(x)$ at the origin. The basic results of Sec. III of BH are contained in the form of the kernel $\hat{K}(x, y)$, see Eq.~(3.21) of BH:
\begin{equation}\label{EQ1}
\hat{K}(x, y) = \frac{\hat{\phi^{\,\prime}}(x) \hat{\psi^{\,\prime}}(y) - \hat{\phi^{\,\prime\prime}}(x) \hat{\psi}(y) - \hat{\phi}(x) \hat{\psi^{\,\prime\prime}}(y)}{x-y},
\end{equation}
where the functions $\hat{\phi}(x)$ and $\hat{\psi}(x)$ are defined by
\begin{equation}\label{eq1}
\hat{\phi}(x) = \frac{1}{\pi} \int_{0}^{\infty} e^{-\ulamek{1}{4} t^{4}} \cos(t x) dt,
\end{equation}
and
\begin{equation}\label{EQ2}
\hat{\psi}(x) = -{\rm Im}\left[\frac{\omega}{\pi} \int_{0}^{\infty} e^{- \ulamek{1}{4} u^{4}}\left(e^{x u \omega} - e^{-x u \omega}\right) du\right],
\end{equation}
where in Eq.~\eqref{EQ2} $\omega = e^{i \ulamek{\pi}{4}}$, see Eqs. (3.7) and (3.13) of BH. Evidently, the functions $\hat{\phi}(x)$ and $\hat{\psi}(x)$ satisfy $\hat{\phi}(-x) = \hat{\phi}(x)$ and $\hat{\psi}(-x) = - \hat{\psi}(x)$. The BH provide, for both $\hat{\phi}(x)$ and $\hat{\psi}(x)$, Taylor expansions for small $x$ and the asymptotic behaviour at large $x$, see \cite{EBrezin98, EBrezin98_1}. These approximations are subsequently used to calculate the density of states given by $\hat{\rho}(x) = \hat{K}(x, x)$ as
\begin{equation}\label{EQ3}
\hat{\rho}(x) = - \left[\hat{\phi}^{\,\prime}(x) \hat{\psi}^{\,\prime\prime}(x) - \hat{\phi}^{\,\prime\prime}(x) \hat{\psi}^{\,\prime}(x) + x \hat{\phi}(x) \hat{\psi}(x)\right],
\end{equation}
as well as the connected correlation function for two eigenvalues symmetric with respect to the origin, given by
\begin{equation}\label{EQ4}
\hat{\rho}_{c}(x, -x) \equiv \hat{\rho}_{c}(x) = - \left[\hat{K}(x, -x)\right]^{2}.
\end{equation}
Here we furnish exact formulas for $\hat{\phi}(x)$ and $\hat{\psi}(x)$ thus enabling to visualize graphically Eqs.~\eqref{EQ3} and \eqref{EQ4} for all the values of parameters. The exact forms of $\hat{\phi}(x)$ and $\hat{\psi}(x)$ are
\begin{equation}\label{EQ5}
\hat{\phi}(x) = \frac{1}{2 \Gamma\big(\ulamek{3}{4}\big)} {_{0}F_{2}}\left({- \atop \ulamek{1}{2}, \ulamek{3}{4}} \Big| \frac{x^{4}}{64}\right) - \frac{\sqrt{2} \Gamma\big(\ulamek{3}{4}\big)}{4 \pi} x^{2} {_{0}F_{2}}\left({- \atop \ulamek{5}{4}, \ulamek{3}{2}} \Big| \frac{x^{4}}{64}\right),
\end{equation}
and 
\begin{equation}\label{EQ6}
\hat{\psi}(x) = -\frac{x}{\sqrt{\pi}} {_{0}F_{2}}\left({- \atop \ulamek{3}{4}, \ulamek{5}{4}} \Big\vert -\frac{x^{4}}{64}\right).
\end{equation}
In Eqs.~\eqref{EQ5} and \eqref{EQ6} ${_{0}F_{2}}$ is the generalized hypergeometric function of type ${_{p}F_{q}}\left(^{(\alpha_{p})}_{(\beta_{q})} \Big| z\right)$ as defined, for example in \cite{APPrudnikov98_3}, with $(\alpha_{p}) = \alpha_{1}, \alpha_{2}, \ldots, \alpha_{p}$ etc. See Appendix A for definition of ${_{p}F_{q}}$.

In Sec. V of BH more involved conditions on $H_{0}$ are chosen and this yields modified form of functions $\phi(x)$ and $\psi(x)$ given by Eqs.~(5.7) and (5.8) of BH respectively as
\begin{equation}\label{EQ7}
\phi(x) = \frac{1}{2 \pi} \int_{-\infty}^{\infty} e^{-\ulamek{t^6}{3} + i t x} dt,
\end{equation}
and 
\begin{equation}\label{EQ8}
\psi(x) = \frac{1}{2 \pi i} \int_{C} e^{-\ulamek{t^6}{3} + t x} dt,
\end{equation}
where in Eq.~\eqref{EQ8} the integration path $C$ consists of four lines of steepest descent starting from the origin to infinity. On these lines $t$ is replaced by $t = e^{\pm i \ulamek{\pi}{3}} t^{\prime}$ and $t = e^{\pm i \ulamek{2 \pi}{3}} t^{\prime}$, respectively. We have evaluated the integrals in Eqs.~\eqref{EQ7} and  \eqref{EQ8} exactly and their explicit forms will be reported in the part \textbf{B} of our Sec. II below. 

The paper is structured as follows: in Sec. 2 we derive the main results for the kernel $\hat{K}(x, y)$ and for its relative $K(x, y)$ built on $\phi(x)$ and $\psi(x)$, and summarize the asymptotics of $\hat{\phi}(x)$, $\hat{\psi}(x)$, $\phi(x)$ and $\psi(x)$. In Sec. 3 we give graphical representations of these functions, of the both kernels and of their associated quantities $\hat{\rho}(x)$, $\rho(x)$, $\hat{\rho}_{c}(x)$ and $\rho_{c}(x)$. We compare also graphically the exact results and the asymptotics of Sec. 3. The Sec. 4 contains discussion and conclusions. Some informations concerning generalized hypergeometric functions are given in the Appendix A. In the Appendix B the definition of Meijer G functions along with the description of relevant complex integration paths are given.

\section{Derivation of the main results}

\noindent
\textbf{A.}$\;\;$ We proceed to a streamlined derivation of the formulas \eqref{EQ5} and \eqref{EQ6}. Thus we refer here to Sec. III of BH. The method is borrowed from our recent exposition of exact and explicit forms of one-sided \cite{KAPenson10} and two-sided \cite{KGorska11, KGorska12} L\'{e}vy stable probability distribution functions. It is based on a combined use of Mellin and Laplace transforms (for one-sided case) and of Mellin and Fourier transforms (for two-sided case). It translates directly to the case when the sought for functions are still of L\'{e}vy-stable character but are not anymore probability distribution functions. According to \cite{PLevy23} here the condition of positivity is not fulfilled and $\hat{\phi}(x)$ and $\hat{\psi}(x)$ are \textit{signed} oscillatory functions. Utility of these functions in a number of applications has been strongly emphasized in \cite{TMGaroni02, TMGaroni08}. 

We define a complex function 
\begin{equation}\label{EQ9}
g_{4}(x) = \frac{1}{2 \pi} \int_{-\infty}^{\infty} e^{-\ulamek{y^{4}}{4}} e^{-i x y} dy.
\end{equation}
It is clear that ${\rm Re}[g_{4}(x)] = \hat{\phi}(x)$ and ${\rm Im}[g_{4}(x)]$ is related to $\hat{\psi}(x)$. We define the Mellin transform of a function $f(x)$, for $s$ complex, as
\begin{equation}\nonumber
\mathcal{M}[f(x); s] = f^{\star}(s) = \int_{0}^{\infty} x^{s-1} f(x) ds,
\end{equation}
and its inverse as
\begin{equation}\label{EQ10}
\mathcal{M}^{-1}[f^{\star}(s); x] = f(x) = \frac{1}{2 \pi i} \int_{L} x^{-s} f^{\star}(s) ds,
\end{equation}
see \cite{INSneddon72} for the appropriate choice of integration contour $L$ in Eq.~\eqref{EQ10}.

In what follows we calculate the Mellin transform of $g_{4}(x)$ from Eq.~\eqref{EQ9}:
\begin{eqnarray}
g_{4}^{\star}(s) &=& \int_{0}^{\infty} x^{s-1} \left[\frac{1}{\pi} \int_{0}^{\infty} e^{-\ulamek{y^{4}}{4}} e^{- i x y} dy\right] dx = \frac{1}{\pi} \int_{0}^{\infty} e^{-\ulamek{y^{4}}{4}} \left[\int_{0}^{\infty} x^{s-1} e^{- i x y} dx\right] dy \label{EQ11} \\[0.7\baselineskip]
&=& \frac{1}{4 \pi} 4^{\ulamek{1-s}{4}}  e^{-i \ulamek{\pi s}{2}} \Gamma(s) \Gamma\left(\ulamek{1-s}{4}\right),  \label{EQ12}
\end{eqnarray}
where $0<{\rm Re}(s)<1$. In Eq.~\eqref{EQ11} we have used the definition of gamma function along with the formulas 2.3.3.1 and 2.3.18.2 in \cite{APPrudnikov98_1}. We employ now in ${\rm Re}[g_{4}^{\star}(s)]$ the second Euler's reflection formula $\Gamma\big(\ulamek{1}{2}+z\big)\Gamma\big(\ulamek{1}{2}-z\big)~=~\ulamek{\pi}{\cos(\pi z)}$, which leads, via Eq.~\eqref{EQ10}, to
\begin{equation}\label{EQ13}
{\rm Re}[g_{4}^{\star}(s)] = \hat{\phi}(x) = \frac{1}{2 \pi i} \int_{L^{'}} x^{-s} 4^{\ulamek{1-s}{4}} \frac{\Gamma(s-1) \Gamma\big(1 - \ulamek{s-1}{4}\big)}{\Gamma\big(1 - \ulamek{s-1}{2}\big) \Gamma\big(\ulamek{s-1}{2}\big)} ds.
\end{equation}
After changing variable $u = \ulamek{s-1}{4}$, Eq.~\eqref{EQ13} becomes
\begin{equation}\label{EQ14}
\hat{\phi}(x) = \sqrt{\frac{2}{\pi}} \frac{1}{x} \frac{1}{2 \pi i} \int_{\tilde{L}^{'}} \left(\frac{x^{4}}{64}\right)^{\!\!-u} \frac{\Gamma\big(\ulamek{1}{4} + u\big) \Gamma\big(\ulamek{3}{4} + u\big)}{\Gamma\big(\ulamek{1}{2} - u\big)} du = \sqrt{\frac{2}{\pi}} \frac{1}{x}\, G^{2, 0}_{0, 3}\left(\frac{x^{4}}{64}\Big| {- \atop \ulamek{1}{4}, \ulamek{3}{4}, \ulamek{1}{2}}\right), 
\end{equation}
where in obtaining Eq.~\eqref{EQ14} we have used in the integrand of Eq.~\eqref{EQ13} three times the Gauss-Legendre multiplication formula for gamma functions. The form of Eq.~\eqref{EQ14} permits to naturally encode it as a Meijer's G function \cite{APPrudnikov98_3}, for which we use the conversion formula 16.17.2 of \cite{NIST} in terms of generalized hypergeometric functions ${_{p}F_{q}}$ (see also \cite{YLLuke69}) and the final expression in the form of Eq.~\eqref{EQ5} results. We shall supply the classical Meijer G notation also for the remaining integrals of similar type i.e. Eqs. \eqref{EQ16}, \eqref{EQ19} and \eqref{EQ20}. In order to meaningfully define the Meijer's G function from Eqs.~\eqref{EQ13} and \eqref{EQ14} the integration contours must be specified. This applies also to the contours in Eqs.~\eqref{EQ16}, \eqref{EQ19} and \eqref{EQ20} below. Consult \cite{APPrudnikov98_3, YLLuke69} for a clear exposition of this procedure. See Appendix B for more details.

\noindent
The calculation of $\hat{\psi}(x)$ follows a similar pattern.

In ${\rm Im}[g_{4}^{\star}(s)]$ we apply the first Euler's reflexion formula $\Gamma(z) \Gamma(1-z) = \ulamek{\pi}{\sin(\pi z)}$ and it yields, after the Mellin inversion of Eq.~\eqref{EQ10}
\begin{equation}\label{EQ15}
{\rm Im}[g_{4}(y)] = \frac{1}{2 \pi i} \int_{L^{''}} y^{-s} 4^{\ulamek{1-s}{4}} \frac{\Gamma(s) \Gamma\big(\ulamek{1-s}{4}\big)}{4 \Gamma\big(\ulamek{s}{2}\big) \Gamma\big(1-\ulamek{s}{2}\big)} ds.
\end{equation}
Eq.~\eqref{EQ15} will be further transformed with the change of variable $u = \ulamek{s-1}{4}$ and subsequent three-fold application of Gauss-Legendre multiplication formula in the integrand of Eq.~\eqref{EQ15}, to arrive at 
\begin{equation}\label{EQ16}
{\rm Im}[g_{4}(y)] = \sqrt{\frac{2}{\pi}} \frac{1}{y} \frac{1}{2 \pi i} \int_{\tilde{L}^{''}} \left(\frac{y^{4}}{64}\right)^{\!\!-u} \frac{\Gamma(-u) \Gamma\big(\ulamek{1}{2} + u\big) \Gamma(1+ u)}{\Gamma\big(\ulamek{1}{4} - u\big) \Gamma\big(\ulamek{3}{4} - u\big)} du = \sqrt{\frac{2}{\pi}} \frac{1}{y} G^{2, 1}_{1, 4}\left(\frac{y^{4}}{64} \Big\vert {1 \atop \ulamek{1}{2}, 1, \ulamek{1}{4}, \ulamek{3}{4}}\right).
\end{equation}
For explicit form of the integration contour $\tilde{L}^{''}$ in Eq. \eqref{EQ16} see Appendix B. We apply anew the conversion formula 16.17.2 of \cite{NIST} to it and this yields:
\begin{equation}\label{EQ17}
{\rm Im}[g_{4}(y)] = \frac{y}{2 \sqrt{\pi}} {_{0}F_{2}}\left({- \atop \ulamek{3}{4}, \ulamek{5}{4}} \Big| \frac{y^{4}}{64}\right) - \frac{y^{3}}{6 \pi} {_{1}F_{3}}\left({1 \atop \ulamek{5}{4}, \ulamek{3}{2}, \ulamek{7}{4}} \Big| \frac{y^{4}}{64}\right).
\end{equation}
The function $\hat{\psi}(x)$ will now be related to Eq.~\eqref{EQ17}. To this end we rewrite Eq.~\eqref{EQ2} using $\sinh(z) = - e^{i\ulamek{\pi}{2}} \sin\big(e^{i\ulamek{\pi}{2}} z\big)$ as
\begin{equation}\nonumber
\hat{\psi}(x) = {\rm Im}\left[\frac{2\tilde{\omega}}{\pi} \int_{0}^{\infty} e^{-\ulamek{u^{4}}{4}} \sin(\tilde{\omega} x u) du\right], \quad \tilde{\omega} = e^{i \ulamek{3 \pi}{4}},
\end{equation}
which gives, according to Eq.~\eqref{EQ17}
\begin{equation}
\hat{\psi}(x) = 2 {\rm Im} \left[-\frac{i x}{2 \sqrt{\pi}} {_{0}F_{2}}\left({- \atop \ulamek{3}{4}, \ulamek{5}{4}} \Big| - \frac{x^{4}}{64}\right) + \frac{x^{3}}{6\pi} {_{1}F_{3}}\left({1 \atop \ulamek{5}{4}, \ulamek{3}{2}, \ulamek{7}{4}} \Big| - \frac{x^{4}}{64}\right)\right] = - \frac{x}{\sqrt{\pi}} {_{0}F_{2}}\left({- \atop \ulamek{3}{4}, \ulamek{5}{4}} \Big| - \frac{x^{4}}{64}\right). \nonumber
\end{equation}
Thus, Eq.~\eqref{EQ6} is substantiated. 
\ \\

\noindent
\textbf{B.}$\;\;$ We refer now to the Sec.~V of BH and consider in more detail our Eqs.~\eqref{EQ7} and \eqref{EQ8}. 

In close analogy to Eq.~\eqref{EQ9} we define
\begin{equation}\label{EQ18}
g_{6}(x) = \frac{1}{2 \pi} \int_{-\infty}^{\infty} e^{-\ulamek{y^6}{3}} e^{- i x y} dy,
\end{equation}
whose Mellin transform turns out to be equal
\begin{equation}\nonumber
g_{6}^{\star}(s) = \frac{1}{6 \pi} 3^{\ulamek{1-s}{6}} e^{-i\ulamek{s \pi}{2}} \Gamma(s) \Gamma\big(\ulamek{1-s}{6}\big), 
\end{equation}
where $0 < {\rm Re}(s) < 1$. We repeat the reasoning applied to the $g_{4}^{\star}(s)$ in Eq.~\eqref{EQ11} and obtain (compare Eqs.~\eqref{EQ11} and \eqref{EQ12} above):
\begin{eqnarray}\label{EQ19}
\phi(x) &=& {\rm Re}[g_{6}(x)] = \sqrt{\frac{3}{\pi}}\, \frac{1}{x}\, \frac{1}{2 \pi i} \int_{L^{\star}} \left(\frac{3 y^6}{6^6}\right)^{\!\!-z} \frac{\Gamma\big(z + \ulamek{1}{6}\big) \Gamma\big(z + \ulamek{1}{2}\big) \Gamma\big(z + \ulamek{5}{6}\big)}{\Gamma\big(\ulamek{1}{3} - z\big) \Gamma\big(\ulamek{2}{3} - z\big)}\, dz \nonumber\\[0.6\baselineskip]
&=& \sqrt{\frac{3}{\pi}}\, \frac{1}{x}\, G^{3, 0}_{0, 5}\left(\frac{3 x^{6}}{6^{6}} \Big\vert {- \atop \ulamek{1}{6}, \ulamek{1}{2}, \ulamek{5}{6}, \ulamek{1}{3}, \ulamek{2}{3}}\right)\nonumber\\[0.6\baselineskip]
&=& \frac{3^{\ulamek{1}{6}}}{3 \Gamma\big(\ulamek{5}{6}\big)} {_{0}F_{4}}\left({- \atop \ulamek{1}{3}, \ulamek{1}{2}, \ulamek{2}{3}, \ulamek{5}{6}} \Big\vert -\frac{3 x^6}{6^6}\right) - \frac{\sqrt{3}}{12\sqrt{\pi}}\, x^2\, {_{0}F_{4}}\left({- \atop \ulamek{2}{3}, \ulamek{5}{6}, \ulamek{7}{6}, \ulamek{4}{3}} \Big\vert -\frac{3 x^6}{6^6}\right) \nonumber\\[0.6\baselineskip]
&+& \frac{3^{\ulamek{5}{6}} \Gamma\big(\ulamek{5}{6}\big)}{144 \pi}\, x^4\, {_{0}F_{4}}\left({- \atop \ulamek{7}{6}, \ulamek{4}{3}, \ulamek{3}{2}, \ulamek{5}{3}} \Big\vert -\frac{3 x^6}{6^6}\right).
\end{eqnarray}
Likewise,
\begin{eqnarray}\label{EQ20}
{\rm Im}[g_{6}(x)] &=& \sqrt{\frac{3}{\pi}}\, \frac{1}{x}\, \frac{1}{2 \pi i} \int_{L^{^{\star \star}}} \left(\frac{3 y^6}{6^6}\right)^{\!\!-z}  \frac{\Gamma\big(z + \ulamek{1}{3}\big) \Gamma\big(z + \ulamek{2}{3}\big) \Gamma\big(z + 1\big) \Gamma(-z)}{\Gamma\big(\ulamek{1}{6} - z\big) \Gamma\big(\ulamek{1}{2} - z\big) \Gamma\big(\ulamek{5}{6} - z\big)}\, dz \nonumber\\[0.6\baselineskip]
&=& \sqrt{\frac{3}{\pi}}\, \frac{1}{x}\, G^{3, 1}_{1, 6}\left(\frac{3 x^{6}}{6^{6}} \Big\vert {1 \atop \ulamek{1}{3}, \ulamek{2}{3}, 1, \ulamek{1}{6}, \ulamek{1}{2}, \ulamek{5}{6}}\right)\nonumber\\[0.6\baselineskip]
&=& \frac{3^{\ulamek{5}{6}}}{9 \Gamma\big(\ulamek{2}{3}\big)}\, x\, {_{0}F_{4}}\left({- \atop \ulamek{1}{2}, \ulamek{2}{3}, \ulamek{5}{6}, \ulamek{7}{6}} \Big\vert - \frac{3 x^6}{6^6}\right) - \frac{3^{\ulamek{2}{3}} \Gamma\big(\ulamek{2}{3}\big)}{36 \pi}\, x^{3}\,  {_{0}F_{4}}\left({- \atop \ulamek{5}{6}, \ulamek{7}{6}, \ulamek{4}{3}, \ulamek{3}{2}} \Big\vert - \frac{3 x^6}{6^6}\right) \nonumber\\[0.6\baselineskip]
&+& \frac{x^{5}}{240 \pi}  {_{1}F_{5}}\left({1 \atop \ulamek{7}{6}, \ulamek{4}{3}, \ulamek{3}{2}, \ulamek{5}{3}, \ulamek{11}{6}} \Big\vert - \frac{3 x^6}{6^6}\right).
\end{eqnarray}
In order to represent Eq.~\eqref{EQ8} as an integral over positive axis we sum up four contributions as specified after Eq.~\eqref{EQ8} and end up with
\begin{equation}\nonumber
\psi(x) = -{\rm Im}\left[\frac{2 \omega_{1}}{\pi} \int_{0}^{\infty} e^{- \ulamek{\xi^6}{3}} \sinh( \omega_{1} x \xi) d\xi\right], \quad \omega_{1} = e^{i \ulamek{\pi}{3}},
\end{equation}
that gives with the help of Eq.~\eqref{EQ20} the final result
\begin{equation}\label{EQ21}
\psi(x) = -\frac{3^{\ulamek{1}{3}}}{3 \Gamma\big(\ulamek{2}{3}\big)}\, x\, {_{0}F_{4}}\left({- \atop \ulamek{1}{2}, \ulamek{2}{3}, \ulamek{5}{6}, \ulamek{7}{6}} \Big\vert \frac{3 x^6}{6^6}\right) + \frac{3^{\ulamek{1}{6}} \Gamma\big(\ulamek{2}{3}\big)}{12 \pi}\, x^{3}\,  {_{0}F_{4}}\left({- \atop \ulamek{5}{6}, \ulamek{7}{6}, \ulamek{4}{3}, \ulamek{3}{2}} \Big\vert \frac{3 x^6}{6^6}\right).
\end{equation}
Thus, Eq.~(5.6) of BH which is (we have removed the "hat" to be consistent with Eqs.~\eqref{EQ7} and \eqref{EQ8})
\begin{equation}
K(x, y) = \frac{2}{x-y} \left[\phi^{(4)}(x) \psi(y) - \phi^{(3)}(x) \psi^{\,\prime}(y) + \phi^{\, \prime\prime}(x) \psi^{\, \prime\prime}(y)  - \phi^{\, \prime}(x) \psi^{(3)}(y) + \phi(x) \psi^{(4)}(y) \right] \label{EQ22}
\end{equation}
gives the density of states
\begin{equation}\label{EQ23}
\rho(x) = - x \phi(x) \psi(x) - 2\left[\phi^{(4)}(x) \psi^{\,\prime}(x) - \phi^{\,\prime}(x) \psi^{(4)}(x) + \phi^{\,\prime\prime}(x) \psi^{(3)}(x) - \phi^{(3)}(x) \psi^{\,\prime\prime}(x)\right],
\end{equation}
where the first term is the previous expression stems from the exact differential equation for $\phi(x)$: $\phi^{(5)}(x)=-\ulamek{x}{2} \phi(x)$. With $\phi(x)$ of Eq.~\eqref{EQ19} and $\psi(x)$ of Eq.~\eqref{EQ21} the density of states $\rho(x)$ and the quantity 
\begin{equation}\label{EQ24}
\rho_{c}(x) = - \left[K(x, -x)\right]^{2}
\end{equation}
analogous to Eq.~\eqref{EQ4} can be readily calculated. 
\ \\

\noindent
\textbf{C.}$\;\;$ Although exact forms of $\hat{\phi}(x)$, $\hat{\psi}(x)$, $\phi(x)$ and $\psi(x)$ render possible all sorts of calculations, it is still instructive to consider the asymptotics of these functions. We shall briefly comment on it now. In doing so it is practical to introduce the so-called L\'{e}vy functions of index $\alpha$, $\alpha > 0$, which are two-sided symmetric L\'{e}vy stable distributions denoted in \cite{KGorska11, KGorska12} as $g(\alpha, 0; x)$, and considered there under the condition $0 < \alpha \leq 2$. They are special case of general two-sided L\'{e}vy function $g(\alpha, \beta; x)$, $\beta \neq -\alpha$, $\alpha > 0$. In the present work $\beta$ is set to zero and therefore \cite{KGorska11, TMGaroni02, TMGaroni08}
\begin{equation}\label{eq2}
g(\alpha, 0; x) = \frac{1}{\pi} \int_{0}^{\infty} e^{-t^{\alpha}} \cos(x t) dt, 
\end{equation}
where $\alpha > 0$ and $-\infty < x < \infty$. As mentioned earlier only for $0 < \alpha < 2$ are these functions probability distribution functions (pdf). For $\alpha > 2$, $g(\alpha, 0; x)$ are not pdf's and are \textit{signed} oscillatory functions. One sees rapidly that
\begin{equation}\label{eq3}
\hat{\phi}(x) = 2^{1/2} g(4, 0; 2^{1/2} x),
\end{equation}
and 
\begin{equation}\label{eq4}
\phi(x) = 3^{1/6} g(6, 0; 3^{1/6} x).
\end{equation}
Although initially dismissed as not being the pdf's, the functions $g(\alpha, 0; x)$ reappear in the literature again. A considerable amount of information is contained in \cite{TMGaroni02, TMGaroni08}. In \cite{KGorska11, KGorska12} we have obtained exact and explicit forms of $g(\alpha, 0; x)$ for admissible rational $\alpha$, which include $\alpha = 4$ and $6$. We emphasize that our formulas for $g(\alpha, 0; x)$ furnish exact expressions even when they are not pdf's. We employ now Eq.~(5.18) of \cite{TMGaroni02} to confirm the asymptotic behaviour for $x\to \infty$ in Eq.~(3.10) of BH and (2.15) of \cite{EBrezin98_1}, that is
\begin{equation}\label{EQ25}
\hat{\phi}(x) \sim \sqrt{\frac{2}{3 \pi}}\, x^{-\ulamek{1}{3}} \exp\big(- \ulamek{3}{8} x^{4/3}\big)\, \cos\big(\ulamek{3 \sqrt{3}}{8} x^{4/3} - \ulamek{\pi}{6}\big).
\end{equation}
Eq.~(5.21) of \cite{TMGaroni02} gives the asymptotics of $\phi(x)$ in the form
\begin{equation}\label{EQ26}
\phi(x) \sim \frac{\left(2/x\right)^{\ulamek{2}{5}}}{\sqrt{5 \pi}}\, \left\{\frac{1}{2}\, \exp[-\kappa(x)] + \exp\left[\cos\big(\ulamek{3 \pi}{8}\big) \kappa(x)\right] \cos\left[\sin\big(\ulamek{3 \pi}{8}\big) \kappa(x) - \frac{\pi}{5}\right]\right\} + \ldots
\end{equation}
for $x\to \infty$ with $\kappa(x) = (5/3)\, \big(x/2\big)^{6/5}$. 

Using the Taylor expansion of cosine in ${\rm Re}[g_{6}(x)]$ in Eq.~\eqref{EQ18} leads to the asymptotic behaviour of $\phi(x)$ at $x\to 0$, namely
\begin{equation}\label{EQ27}
\phi(x) \sim \frac{3^{-\ulamek{5}{6}}}{2 \pi} \sum_{r=0}^{\infty} \frac{(-3^{\ulamek{1}{3}})^{r}}{(2 r)!} \Gamma\left(\frac{r}{3} + \frac{1}{6}\right) x^{2 r}.
\end{equation}
The series of  Eq. \eqref{EQ27} can be shown to have an infinite radius of convergence. Its hypergeometric representation is given by Eq. \eqref{EQ19}, and it follows naturally from the Mellin transform method exposed above.

To find the asymptotic behaviour of $\hat{\psi}(x)$ for $x\to \infty$ we have a ready-to-use formula 6.46 on p. 96 of \cite{OIMarichev83}, i.e.
\begin{equation}
\frac{{_{0}F_{n}}\left({- \atop (c_{n})} \Big\vert -z\right)}{\prod_{k=1}^{n} \Gamma(c_{k})} \sim \frac{2 (2 \pi)^{-\ulamek{n}{2}}}{\sqrt{1+n}}\, \exp\left[\frac{z^{\ulamek{1}{1+n}}}{(1+n)^{-1}} \cos\left(\frac{\pi}{1+n}\right)\right] z^{\xi} \cos\left[\pi \xi + \frac{z^{\ulamek{1}{1+n}}}{(1+n)^{-1}} \sin\left(\frac{\pi}{1+n}\right)\right], \nonumber
\end{equation}
which is valid for $n=2, 3, 4, \ldots$, ${\rm arg}(z) = 0$ or for $n=1$, $|{\rm arg}(z)| < 2\pi$, with $\lambda = \sum_{k=1}^{\infty} c_{k}$, $\xi = \frac{n/2 - \lambda}{1+n}$, $z\to \infty$. That gives
\begin{equation}\label{EQ28}
\hat{\psi}(x) \sim \sqrt{\frac{2}{3 \pi}} x^{-\ulamek{1}{3}}\, \exp\big(\ulamek{3}{8} x^{4/3}\big)\, \cos\big(\ulamek{3 \sqrt{3}}{8} x^{4/3} + \ulamek{2 \pi}{3}\big),
\end{equation}
which apart the factor 2, agrees with Eq.~(3.16) of BH, and Eq.~(2.21) of \cite{EBrezin98_1}.\\
By similar means we conjecture the asymptotics of $\psi(x)$, $x \to \infty$, to be
\begin{equation}\label{EQ29}
\psi(x) \sim \frac{\left(2/x\right)^{\ulamek{2}{5}} }{\sqrt{5 \pi}}\, \left\{\frac{1}{2}\, \exp[-\kappa(x)] - \exp\left[-\cos\big(\ulamek{3 \pi}{8}\big) \kappa(x)\right] \sin\left[\sin\big(\ulamek{3 \pi}{8}\big) \kappa(x) + \frac{\pi}{5}\right]\right\} + \ldots \,.
\end{equation}
The asymptotic behaviour of $\phi(x)$ at $x\to 0$ can be obtained by taking Taylor expansion of Eq.~\eqref{EQ21}. That yields to
\begin{equation}\label{EQ30}
\psi(x) \sim -\frac{3^{\ulamek{1}{3}}}{3 \Gamma\big(\ulamek{2}{3}\big)} x + \frac{3^{\ulamek{1}{6}} \Gamma\big(\ulamek{2}{3}\big)}{12 \pi} x^{3} - \frac{3^{\ulamek{4}{3}}}{6^{6} \Gamma\big(\ulamek{2}{3}\big)} x^{7} + \ldots
\end{equation}

Substituting the asymptotics at $x\to\infty$ of $\hat{\phi}(x)$, $\hat{\psi}(x)$ and $\phi(x)$, $\psi(x)$ it is easy to estimate the behaviour of $\hat{\rho}(x)$ (see Eq.~\eqref{EQ3}) and $\rho(x)$ (see Eq.~\eqref{EQ23}) for large $x$. That gives the expressions whose first terms with the leading powers are proportional to $x^{1/3}$ and $x^{1/5}$, for $\hat{\rho}(x)$ and $\rho(x)$, respectively. 

In the following Section we shall compare this asymptotics with the exact results.

\section{Graphical Representation}

\begin{figure}[!h]
\begin{center}
\includegraphics[scale=0.4]{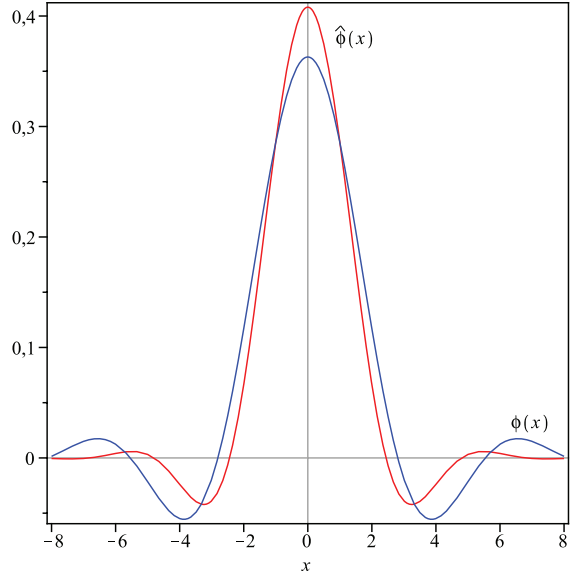}
\caption{\label{fig1} (Color online) Plot of the functions $\hat{\phi}(x)$, see Eq.~\eqref{EQ5}, red line, and $\phi(x)$, see Eq.~\eqref{EQ19}, blue line.}
\end{center}
\end{figure}

\begin{figure}[!h]
\begin{center}
\includegraphics[scale=0.4]{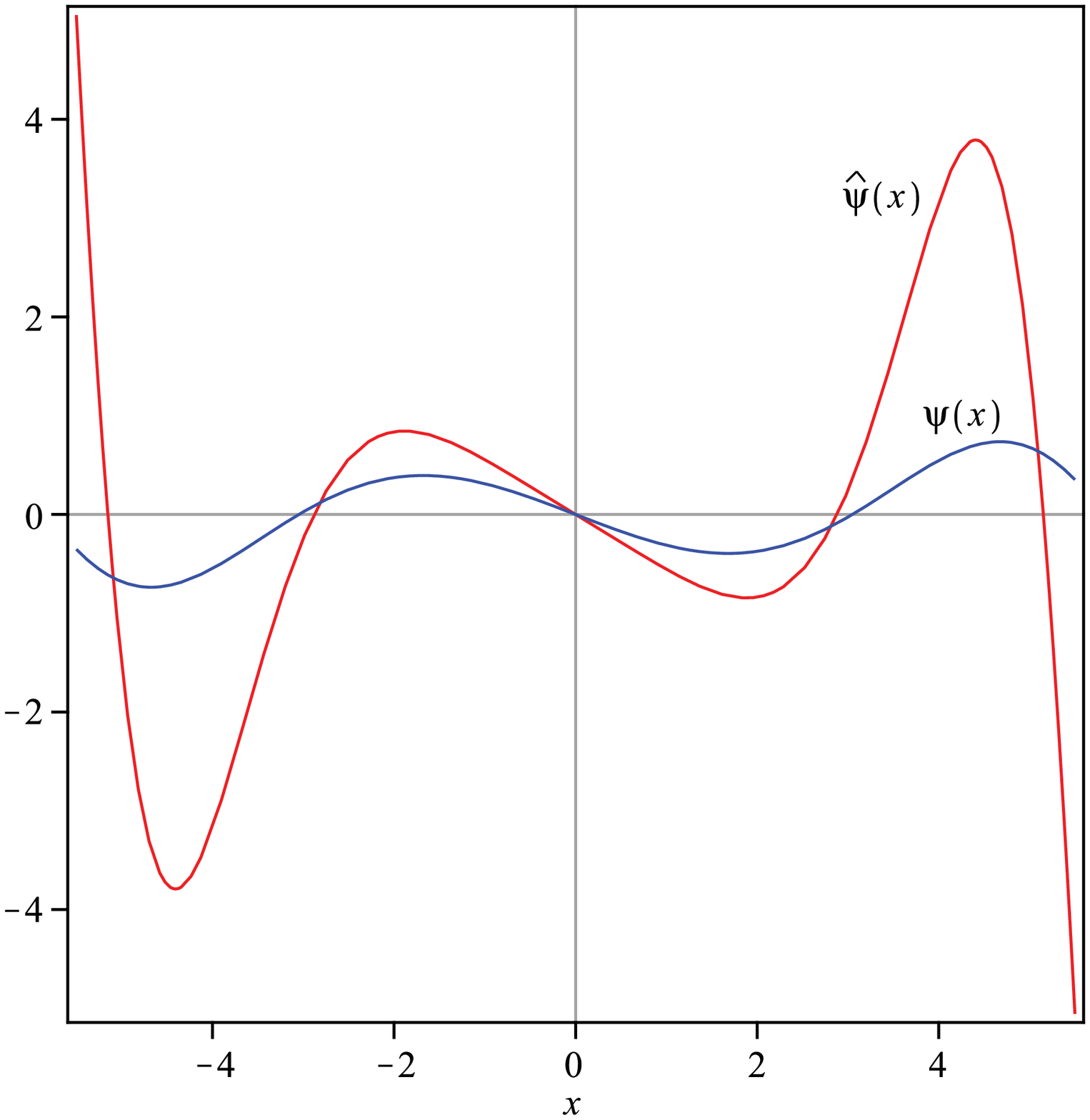}
\caption{\label{fig2} (Color online) Plot of the functions $\hat{\psi}(x)$, see Eq.~\eqref{EQ6}, red line, and $\psi(x)$, see Eq.~\eqref{EQ21}, blue line.}
\end{center}
\end{figure}

\begin{figure}[!h]
\begin{center}
\includegraphics[scale=0.4]{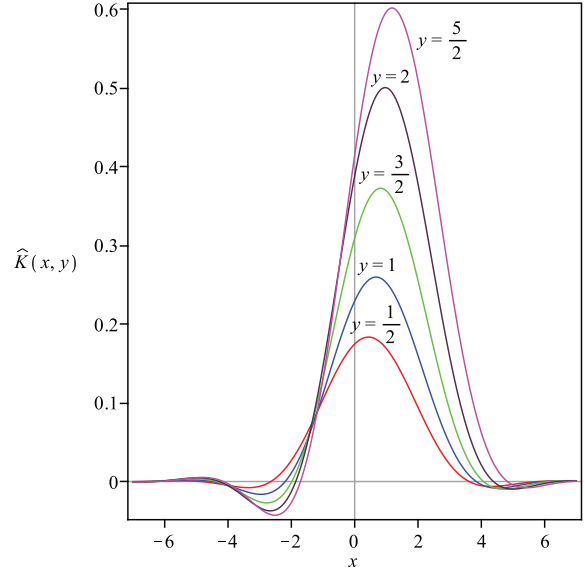}
\caption{\label{fig3} (Color online) Some sections through the surface $\hat{K}(x, y)$, see Eq.~\eqref{EQ1}, as a function of $x$, for: $y = 1/2$ (red line), $y=1$ (blue line), $y = 3/2$ (green line), $y=2$ (violet line), $y=5/2$ (magenta line).}
\end{center}
\end{figure}

As the formulas from Secs.~2, \textbf{A} and 2, \textbf{B} are valid for arbitrary $x$, we have carried out graphical presentation of selected quantities. In Fig. \ref{fig1} we display the oscillating functions $\hat{\phi}(x)$ and $\phi(x)$, see Eq.~\eqref{EQ5} and \eqref{EQ19} respectively. The amplitude of these oscillators dies down very rapidly with $x$. In Fig. \ref{fig2} we present the oscillating functions $\hat{\psi}(x)$ and $\psi(x)$, see Eq.~\eqref{EQ6} and \eqref{EQ21}, respectively. Here, the amplitudes increase drastically as a functions of $x$. In Fig. \ref{fig3} some sections through the surface $\hat{K}(x, y)$ (see Eq.~\eqref{EQ1}) are presented, for fixed values of $y$, as a function of $x$. We have also obtained similar curves for $K(x, y)$ from Eq.~\eqref{EQ22} but we shall not present them here. 

\begin{figure}[!h]
\begin{center}
\includegraphics[scale=0.4]{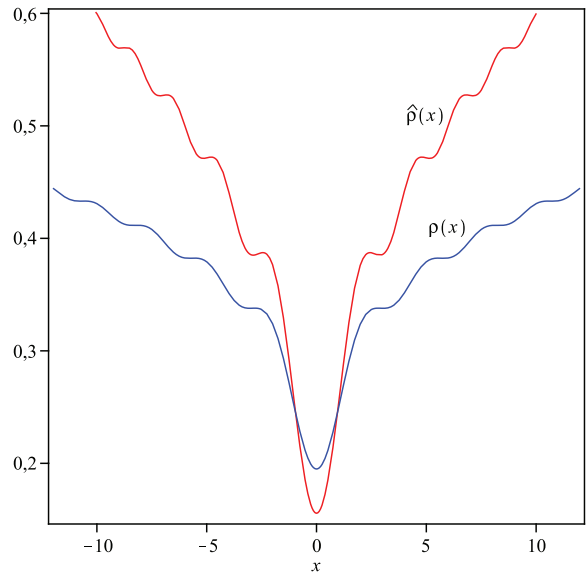}
\caption{\label{fig4} (Color online) Plot of densities of states: $\hat{\rho}(x)$, see Eq.~\eqref{EQ3}, red line, and $\rho(x)$, see Eq.~\eqref{EQ23}, blue line.}
\end{center}
\end{figure}

\begin{figure}[!h]
\begin{center}
\includegraphics[scale=0.4]{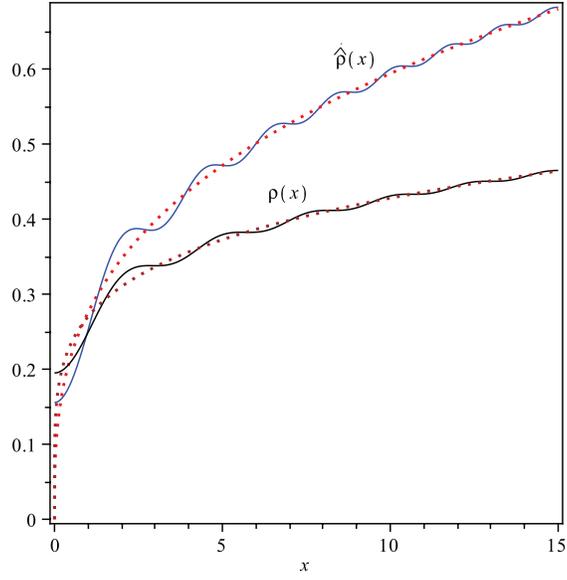}
\caption{\label{fig5} (Color online) Plot of densities of states: $\hat{\rho}(x)$ see Eq.~\eqref{EQ3}, blue line, and its large-$x$ asymptotics, $\sim 0.276 x^{1/3}$, dotted red line; $\rho(x)$ see Eq.~\eqref{EQ23}, black line and its large-$x$ asymptotics, $\sim 0.270 x^{1/5}$, dotted brown line. For transparency the dotted curves represent the leading power approximations, without the inclusion of the oscillating terms.}
\end{center}
\end{figure}

\begin{figure}[!h]
\begin{center}
\includegraphics[scale=0.4]{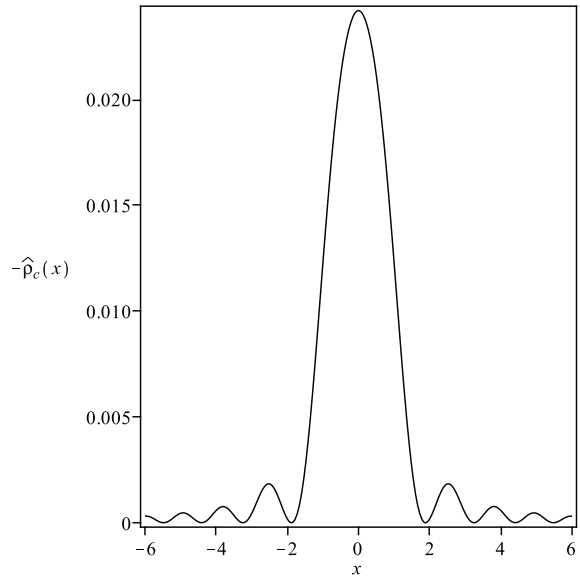}
\caption{\label{fig6} Plot of correlation function $-\hat{\rho}_{c}(x)$, see Eq.~\eqref{EQ4}.}
\end{center}
\end{figure}

\begin{figure}[!h]
\begin{center}
\includegraphics[scale=0.4]{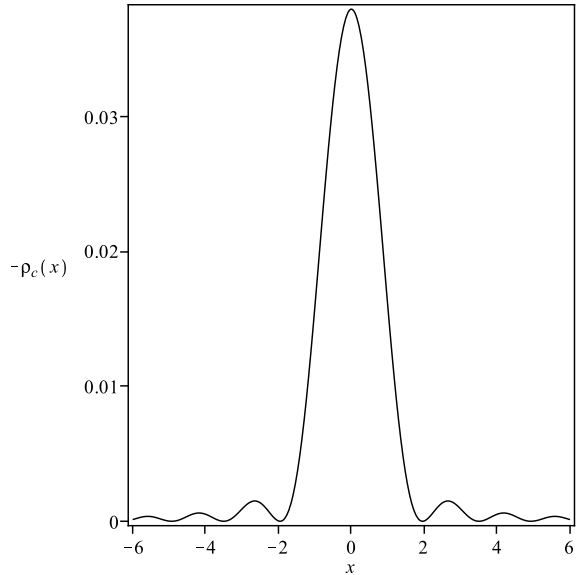}
\caption{\label{fig7} Plot of correlation function $-\rho_{c}(x)$, see Eq.~\eqref{EQ24}.}
\end{center}
\end{figure}

The advantage of our approach is the possibility to analyze the physical quantities in the whole range of their parameters. We carry it out here for the key quantity of the present theory, the exact density of states $\rho(x)$. In Fig. \ref{fig4} we have presented the densities of states $\hat{\rho}(x)$ and $\rho(x)$ from Eq.~\eqref{EQ3} and \eqref{EQ23} respectively. They present a non-monotonic behaviour with characteristic pockets where local minima appear. This neatly complements the results of BE. In Fig. \ref{fig5} the same quantities are compared with their asymptotics for large $x$, see Sec. 2, \textbf{C}. For transparency, in Fig. \ref{fig5} we have retained for the dotted curves only the leading power-law terms. The oscillatory contributions to the asymptotics are not displayed. We stress however that the oscillations in the exact solutions (see continuous curves) agree very well with the oscillations given by asymptotics, especially for $x>20$. In Fig. \ref{fig6} the correlation function $-\hat{\rho}_{c}(x)$ is displayed, see Eq.~\eqref{EQ4}. In Fig. \ref{fig7} we illustrate the correlation function $-\rho_{c}(x)$, see Eq.~\eqref{EQ24}.

\section{Discussion and Conclusions}

The key results of present investigation are contained in the exact expressions for the kernel $\hat{K}(x, y)$ in Eq.~\eqref{EQ1} and for the kernel $K(x, y)$ in Eq.~\eqref{EQ22}. This allows one to obtain with an ease the density of states $\hat{\rho}(x)$ and $\rho(x)$ which display a characteristic oscillatory non-monotonic behaviour, with the infinite series of local minima. 

It is also instructive to compare on the same plot the exact expressions for the kernels with their approximations obtained by numerical evaluation of integrals of Eqs. \eqref{eq1} and \eqref{EQ2} for a given precision $p$, the number of significative digits retained. Since the surfaces given by $\hat{K}(x, y)$ and $K(x, y)$ are in general rich, we shall limit ourselves to consider for illustrative purpose only two sections of the surface $\hat{K}(x, y)$. Clearly, other sections can be similarly analyzed. 

\begin{figure}[!h]
\begin{center}
\includegraphics[scale=0.4]{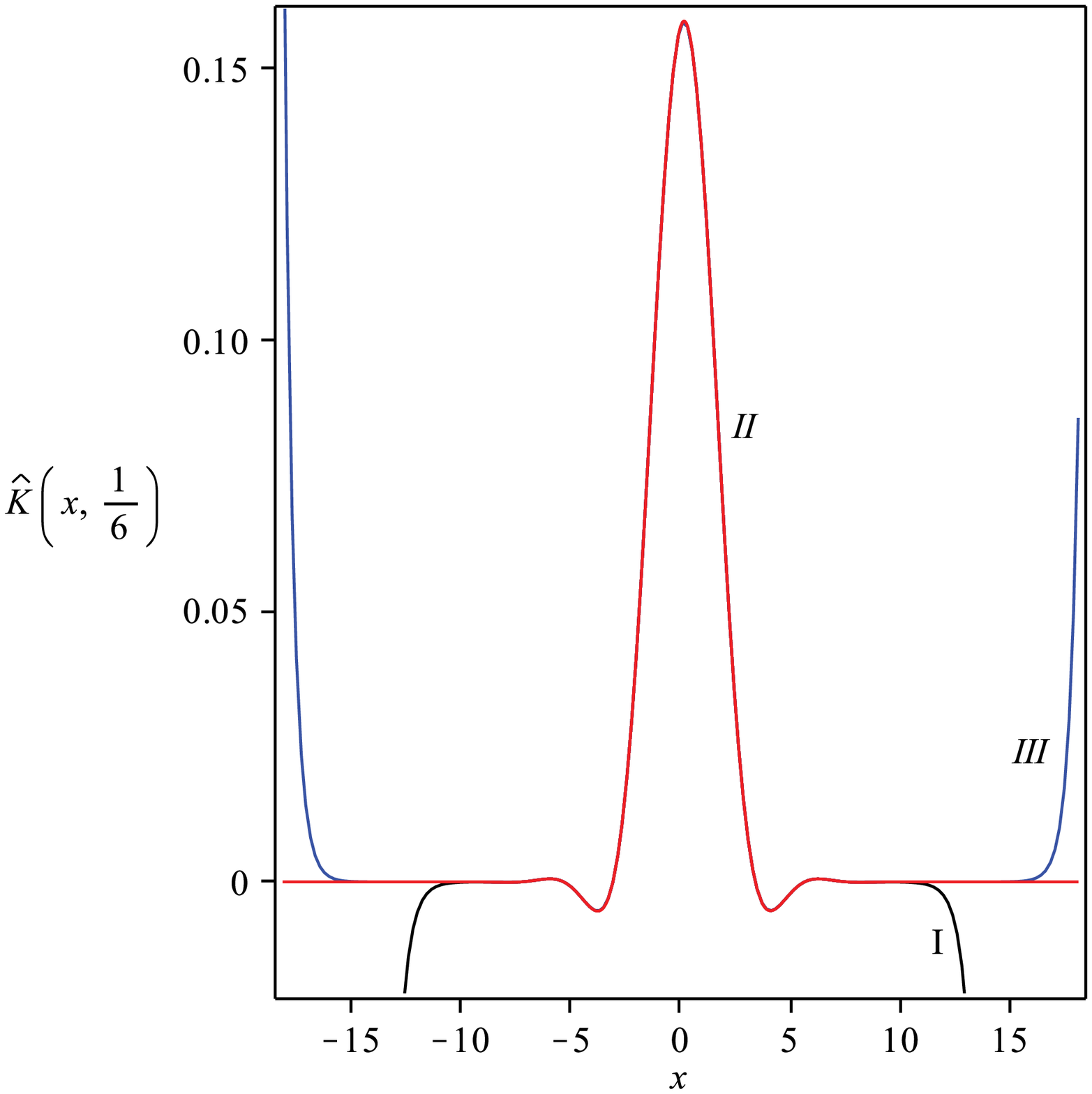}
\caption{\label{fig8} (Color online) Comparison of exact values of $\hat{K}\big(x, \ulamek{1}{6}\big)$, (red curve, II) with the numerical approximations with $p=10$, (black curve, I) and $p=15$, (blue curve, III).}
\end{center}
\end{figure}

\begin{figure}[!h]
\begin{center}
\includegraphics[scale=0.4]{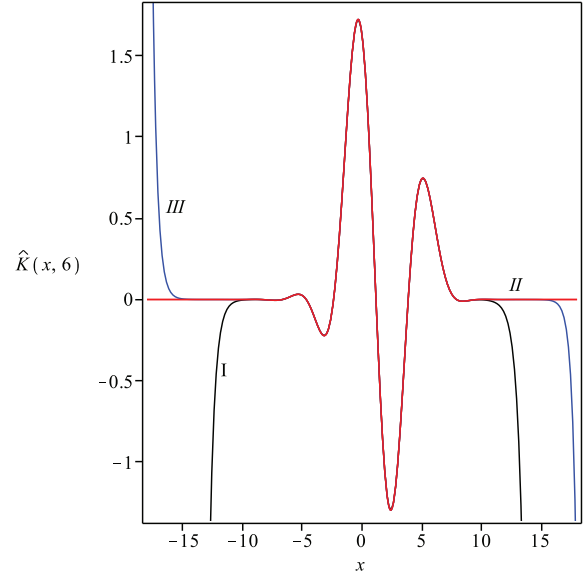}
\caption{\label{fig9} (Color online) Comparison of exact values of $\hat{K}(x, 6)$, (red curve, II) with the numerical approximations with $p=10$, (black curve, I) and $p=15$, (blue curve, III).}
\end{center}
\end{figure}

In Fig. \ref{fig8} we compare the exact values of $\hat{K}\big(x, \ulamek{1}{6}\big)$ (red curve, II) with the numerical approximation with the precision $p=10$ (black curve, I) and with the precision $p=15$ (blue curve, III). As expected, the regions of agreement increase with the imposed numerical precision. Note the appearance of spurious divergencies in the numerical results beyond the range of agreement. In Fig. \ref{fig9} we present the analogous comparison for $\hat{K}(x, 6)$ where the exact results (red curve, II) are compared with the approximations for $p=10$ (black curve, I) and $p=15$ (blue curve, III). These results indicate that the care must be exercised when using the approximations and that the preference should be given to the exact results available through Eqs. \eqref{EQ5} and \eqref{EQ6}. We want to stress that the unified hypergeometric encoding used throughout this work presents clear calculational advantages, as these functions are very well implemented in current computer algebra system. In the contrary, the calculations with hand-defined series often lead to convergence problems, as our experience indicates, see \cite{KAPenson10, KGorska11}.

One observes that both pairs of relevant functions $\hat{\phi}(x)$, $\hat{\psi}(x)$ and $\phi(x)$, $\psi(x)$ can be expressed with a particular subset of generalized hypergeometric functions, namely the type ${_{0}F_{n}}\left({-\atop (c_{n})} \big| x\right)$, $n = 1, 2, \ldots$, where $c_{n}$ are positive parameters. These functions have been investigated in the past \cite{OIMarichev83, VKiryakova97} and are sometimes referred to as hyper-Bessel functions \cite{OIMarichev83}. They play a prominent role in the fractional calculus \cite{VKiryakova94} which in turn is a basic tool to investigate non-standard diffusion in statistical mechanics \cite{KGorska12-1}.

The functions encountered in the course of our evaluations can be shown to be related to oscillatory integrals used to describe wave propagation and optical diffraction. In particular, our function $g_{4}(x)$ is closely related to the so-called Pearcey's integral \cite{RBParis01}. Its various forms and asymptotics can be found in Chap. 36 of \cite{NIST}. We hope that the method presented in this work can be also applied to related problems of QCD \cite{MNowak}.

\newpage

We believe that the functions of L\'{e}vy type will be also appearing in further studies of spectra of random matrices.

We thank Professor M. A. Nowak for kindly informing us about Ref. \cite{MNowak}.

The authors have been supported by the Program PHYSCOMB no. ANR-08-BLAN-0243-2 of Agence Nationale de la Recherche (Paris, France) and by the PHC Polonium, Campus France, project no. 288372A.

\appendix
\section{Generalized hypergeometric function ${_{p}F_{q}}\left({(a_{p}) \atop (b_{q})}\Big| z\right)$}

We first define the list of parameters $(a_{p}) = a_{1}, a_{2}, \ldots, a_{p}$ and similarly $(b_{q}) = b_{1}, b_{2}, \ldots, b_{q}$. The Pochhammer symbol $(a)_{k}$ is defined as $(a)_{k} = \Gamma(a+k)/\Gamma(a)$, where $\Gamma(z)$ is Euler's gamma function. The generalized hypergeometric function of type $(p, q)$, $p, q = 0, 1, 2, \ldots$ is defined as the following series \cite{APPrudnikov98_3}:
\begin{equation}\label{A1}
{_{p}F_{q}}\left({(a_{p}) \atop (b_{q})} \Big| z\right) = \sum_{k=0}^{\infty} \frac{(a_{1})_{k} (a_{2})_{k}\ldots (a_{p})_{k}}{(b_{1})_{k} (b_{2})_{k}\ldots (b_{q})_{k}} \frac{z^{k}}{k!},
\end{equation}
where $b_{j} \neq 0, -1, -2, \ldots$, $j = 1, 2, \ldots, q$. The quantities $a_{i}$ are called the ''upper'' parameters and $b_{j}$ are called the ''lower'' parameters. If $p \leq q$, the series of Eq. \eqref{A1} has an infinite radius of convergence, $|z| < \infty$. Thus this applies to our results exemplified in Eqs. \eqref{EQ5}, \eqref{EQ6} and \eqref{EQ21} which involve the functions of type ${_{0}F_{2}}$ and ${_{0}F_{4}}$. The calculation of quantities defined in Eqs. \eqref{EQ1}, \eqref{EQ3}, \eqref{EQ4}, \eqref{EQ22}, \eqref{EQ23} and \eqref{EQ24} require the knowledge of higher derivatives of the functions ${_{0}F_{2}}$ and ${_{0}F_{4}}$. Due to formula 7.2.3.47 of \cite{APPrudnikov98_3}:
\begin{equation}\label{A2}
\frac{d^{n}}{d z^{n}} {_{p}F_{q}}\left({(a_{p}) \atop (b_{q})} \Big| z\right) = \frac{\prod (a_{p})_{n}}{\prod (b_{q})_{n}} {_{p}F_{q}}\left({(a_{p}) +n \atop (b_{q}) + n} \Big| z\right),
\end{equation}
the differentiation of ${_{p}F_{q}}$ essentially gives again the function ${_{p}F_{q}}$ with modified ''upper'' and ''lower'' parameters, and this does not alter the convergence properties. To sum up, all the exact series encountered in this work have infinite radius of convergence. This important circumstance greatly facilitates the calculations.

\section{Integration paths for the Meijer G functions}

In order to establish the integration paths in Eqs. \eqref{EQ14}, \eqref{EQ16}, \eqref{EQ19} and \eqref{EQ20} we should make precise the definition of the Meijer G functions appearing there; these are equal to the following Mellin-Barnes type integral \cite{APPrudnikov98_3, YLLuke69, RBParis01}:
\begin{equation}\label{B1}
G^{m, n}_{p, q}\left(z\Big| {(a_{p}) \atop (b_{q})}\right) = \frac{1}{2\pi i} \int_{L} \frac{\left[\prod_{j=1}^{m} \Gamma(b_{j} + s)\right] \left[\prod_{j=1}^{n} \Gamma(1-a_{j} - s)\right]}{\left[\prod_{j=n+1}^{p} \Gamma(a_{j} + s)\right] \left[\prod_{j=m+1}^{p} \Gamma(1-b_{j} - s)\right]} z^{-s} ds,
\end{equation}
where $z\neq 0$, $0 \leq m \leq q$, $0 \leq n \leq p$, with the complex parameters $a_{j}$, $1 \leq j \leq p$, and $b_{j}$, $1 \leq j \leq q$. If in \eqref{B1} an empty product occurs, it is taken to be equal one. In order to ensure the convergence of \eqref{B1} the infinite contour $L$ must separate the left poles of the numerator given by $s = -b_{j} - k$, $j=1, 2, \ldots, m$, $k=0, 1, 2, \ldots$ from the right poles of the numerator given by $s = 1 - a_{j} + k$, $j=1, 2, \ldots, n$, $k=0, 1, 2, \ldots$. Under specific conditions satisfied by the auxiliary parameters
\begin{align}\label{B2}
c^{\star} &= m+n - \frac{p+q}{2}, \\[0.4\baselineskip]
\mu &= \sum_{j=1}^{q}b_{j} - \sum_{j=1}^{p}a_{j} + \frac{p-q}{2} + 1, \label{B3}
\end{align}
the path $L$ can be of three types: $L_{-\infty}$, $L_{\infty}$ or $L_{i\infty}$, see conditions 8.2.1.1, cases 1)-4), of \cite{APPrudnikov98_3}, where $L_{i\infty}$ can be sometimes reduced to a straight line $(\gamma-i\infty, \gamma+i\infty)$, with $\gamma = \lim_{s\to \infty} {\rm Re}(s)$, $s\in L_{i\infty}$. The evaluation of the parameters $c^{\star}$ and $\mu$ permits one to find the appropriate form of $L$. As an example we shall describe now the contour $\tilde{L}^{''}$ in Eq. \eqref{EQ16}, in which $G^{2, 1}_{1, 4}$ appears. In this case $c^{\star} = \ulamek{1}{2} > 0$ and from case 1) of 8.2.1.1 of \cite{APPrudnikov98_3} the function $G^{2, 1}_{1, 4}$ converges for $|{\rm arg}(z^4/64)| < \ulamek{\pi}{2}$. We have to determine the poles of the numerator of Eq. \eqref{EQ16}. There are two series of such poles: one is constituted by the poles of $\Gamma(s+1/2) \Gamma(s+1)$ and is given by $s = -\ulamek{t}{2}$, $t=1, 2, 3, \ldots$. The second series are the poles of $\Gamma(-s)$ given by $s=0, 1, 2, \ldots$. Set now $-\ulamek{1}{2} < \gamma < 0$. Accordingly, any straight line which starts at the point $\gamma-i\infty$ and finishes at the point $\gamma+i\infty$, will separate the two series of poles of the numerator. Therefore, the sought for contour $\tilde{L}^{''}$ can be any such a straight line with $-\ulamek{1}{2}< \gamma < 0$. Similar consideration can be applied to determine the remaining contours in Eqs. \eqref{EQ14}, \eqref{EQ19} and \eqref{EQ20} and they will not be reproduced here.


\end{document}